\documentclass[12pt,a4paper,fleqn]{article}
\usepackage[textheight=23cm,textwidth=15cm]{geometry}
\usepackage{amsmath}
\usepackage{graphicx}
\usepackage[american]{babel}
\usepackage{here}
\usepackage[journal=jpcafh,articletitle=true,maxauthors=10,etalmode=truncate]{achemso}
\usepackage{caption} 

\begin{document}

\begin{center}

{\LARGE\bf
Immersive Interactive Quantum Mechanics for Teaching and Learning Chemistry
}

\vspace{1cm}

{\large
Thomas Weymuth$^{a,}$\footnote{ORCID: 0000-0001-7102-7022}
and Markus Reiher$^{a,}$\footnote{Corresponding author; e-mail: markus.reiher@phys.chem.ethz.ch; ORCID: 0000-0002-9508-1565}
}\\[4ex]

$^{a}$ Laboratory of Physical Chemistry, ETH Zurich, \\
Vladimir-Prelog-Weg 2, 8093 Zurich, Switzerland

November 06, 2020

\vspace{.41cm}

\end{center}

\begin{center}
\textbf{Abstract}
\end{center}
\vspace*{-.41cm}
{\small

The impossibility of experiencing the molecular world with our senses hampers teaching and understanding
chemistry because very abstract concepts (such as atoms, chemical bonds, molecular structure, reactivity)
are required for this process. Virtual reality, especially when based on explicit physical modeling
(potentially in real time), offers a solution to this dilemma. Chemistry teaching can make use of advanced
technologies such as virtual-reality frameworks and haptic devices. We show how an immersive learning
setting could be applied to help students understand the core concepts of typical chemical reactions by
offering a much more intuitive approach than traditional learning settings. Our setting relies on an
interactive exploration and manipulation of a chemical system; this system is simulated in real-time with
quantum chemical methods, and therefore, behaves in a physically meaningful way.


}

\newpage

\section{Introduction}
\label{sec:introduction}

To understand chemistry, students in schools and at universities are exposed to highly abstract ideas and
concepts about a tiny molecular world that is elusive to our senses. In fact, it is this inaccessibility of 
the molecular world that sets chemistry apart from neighboring disciplines such as physics and biology: we 
understand macroscopic observations of chemical reactions (such as color changes) solely in terms of a dance of atoms
at the nanometer scale in the molecular program.

Chemical concepts are introduced and explained by
following either an abstract mathematical approach or by a sequence of examples. Whereas the former
requires a good understanding of mathematical ideas combined with imaginative capabilities, the latter
approach is hampered by the need of advanced pattern recognition capabilities. A viable
alternative to such an approach might be offered by explicit physical modeling in virtual reality.

Computer visualizations in videos and alike are often used to display complex dynamical processes
like the rearrangements of atoms in a chemical reaction. While there is evidence that such dynamical
visualizations are indeed helpful\cite{Hoffler2007}, other studies have shown that movies are often not more 
effective than static pictures\cite{Tversky2002}. A possible explanation for this observation is that any 
dynamical representation is transient in nature, hence requiring the student to remember previously obtained
information while simultaneously acquiring new information. The resulting high load on the working memory
hinders effective learning\cite{DeKoning2011}. Van Gog \textit{et al.}~argue that especially in cases where
human movements are depicted (\textit{e.g.}, a surgeon carrying out a complicated procedure), students can
benefit a lot by watching a corresponding dynamical representation\cite{VanGog2009}. They argue that in
such cases, the observation of a human performing a certain task automatically activates the so-called mirror
neuron system, which is not activated if the student watches a depiction of a non-human movement. The
mirror neurons might lead to a reduced load on the working memory, thereby explaining the effectiveness
of videos in such cases. Moreover, it is known that the learning effect of all kinds of dynamical 
visualizations, also those depicting non-human motions, are enhanced by involving a student's motor 
system\cite{DeKoning2011}.

\begin{figure}[H]
\begin{center}
\includegraphics[scale=1.0]{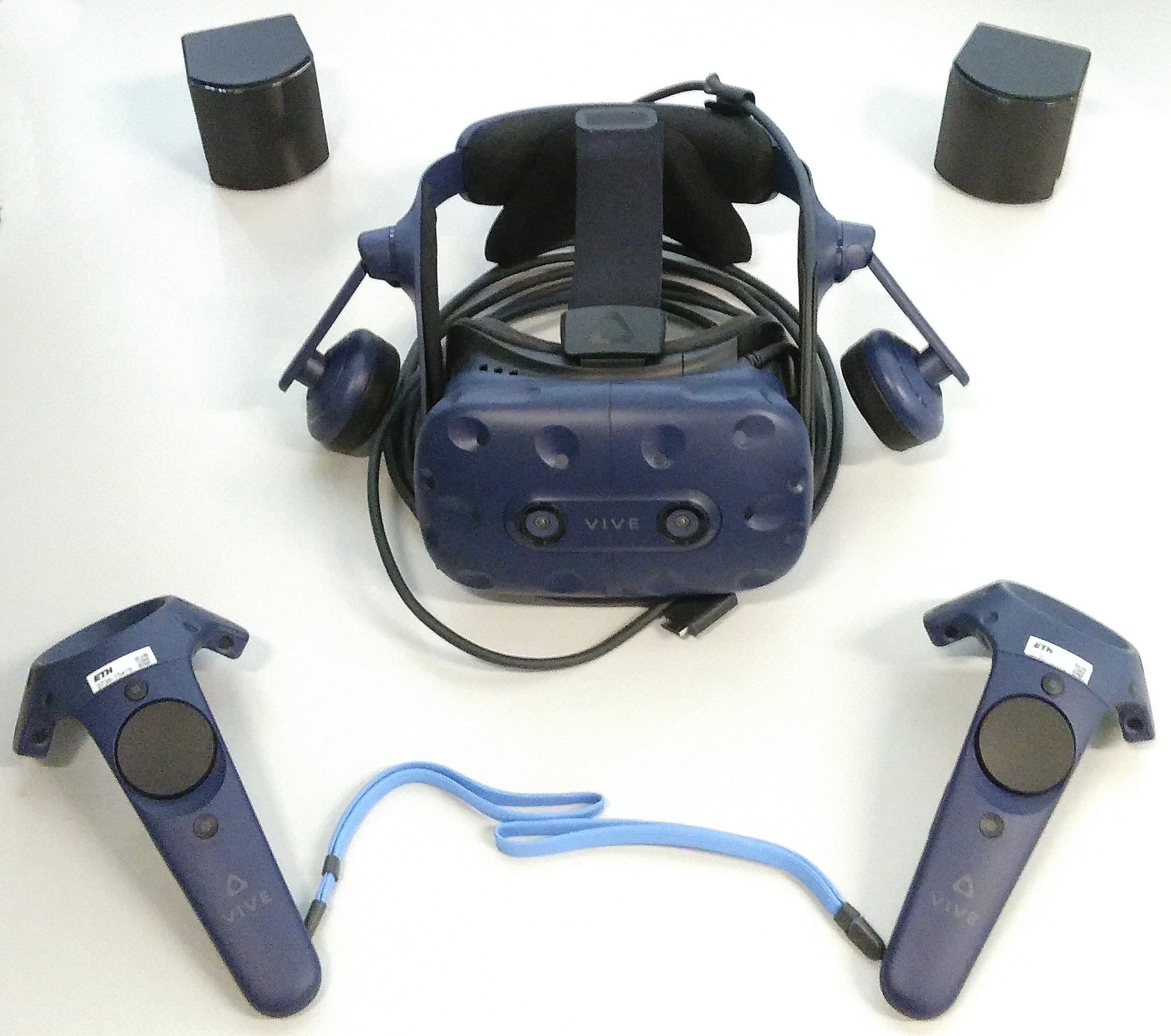}
\end{center}
\caption{\label{fig:virtual_device}\small A virtual reality device (here: ``Vive Pro'' from HTC) consisting of 
a headset (center), two hand-held control sticks (bottom left and right) as well as two base stations used to 
track a persons position (top left and right).}
\end{figure}

In chemistry, technologies such as virtual or augmented reality and haptic devices have been introduced but
have not found widespread application so far\cite{atkinson1977, ouhyoung1989, surles1994, leech1996, 
stone2001, gillet2005, murayama2007, limniou2008, Marti2009, delalande2009, bosson2012, Haag2013, Haag2014, Haag2014a,
faraday2014, glowacki2014, luehr2015, norrby2015, salvadori2016, zheng2017, oconnor2018, muller2018, kingsley2019, 
garciahernandez2019, ferrell2019, dai2020, doak2020, gandhi2020, sanii2020, molegram}. 
Virtual and augmented reality technologies allow a person 
to be immersed into a virtual world. In virtual reality, the entire world experienced by a person is virtual, 
whereas in augmented reality (sometimes also called mixed reality), the person experiences the real world 
while additional virtual objects are added to it, \textit{e.g.}, through a suitable projection on a head-up 
display. Currently, the immersion into the virtual world is most easily achieved with a headset, which 
provides a stereoscopic head-mounted display, stereo sound, as well as head motion tracking sensors (see 
Fig.~\ref{fig:virtual_device}). With this, a person is able to look around in and explore the virtual world. 
For this to be sensible, it is decisive that reliable physical modeling is used. In addition, the movement of 
the entire body of a person is often tracked by means of fixed external sensors. This then allows a person to 
move around freely in the virtual world. Finally, hand-operated controllers are used to interact with the 
objects of the virtual world. Such a fully immersive, virtual reality holds a great potential to help students 
understand better a range of chemical phenomena (see, \textit{e.g.}, Refs.~\citenum{dede1999, trindade2002, 
limniou2008, ferrell2019, bennie2019, fung2019, jimenez2019, dai2020, gandhi2020, ali2020, sanii2020}). For 
example, the structure of a large molecule with a complicated three-dimensional shape can easily be inspected 
from all possible angles, and the rearrangements of the individual atoms of this molecule as it undergoes a 
certain chemical reaction can be examined with great scrutiny\cite{limniou2008, glowacki2014, norrby2015, 
salvadori2016, zheng2017, oconnor2018, muller2018, bennie2019, ferrell2019, jimenez2019, kingsley2019, 
garciahernandez2019, doak2020, gandhi2020, sanii2020}. Since the person experiencing such a virtual reality is 
fully immersed into it, it should provide an ideal environment for teaching, learning, and, of course, 
research.

\begin{figure}[H]
\begin{center}
\includegraphics[scale=1.0]{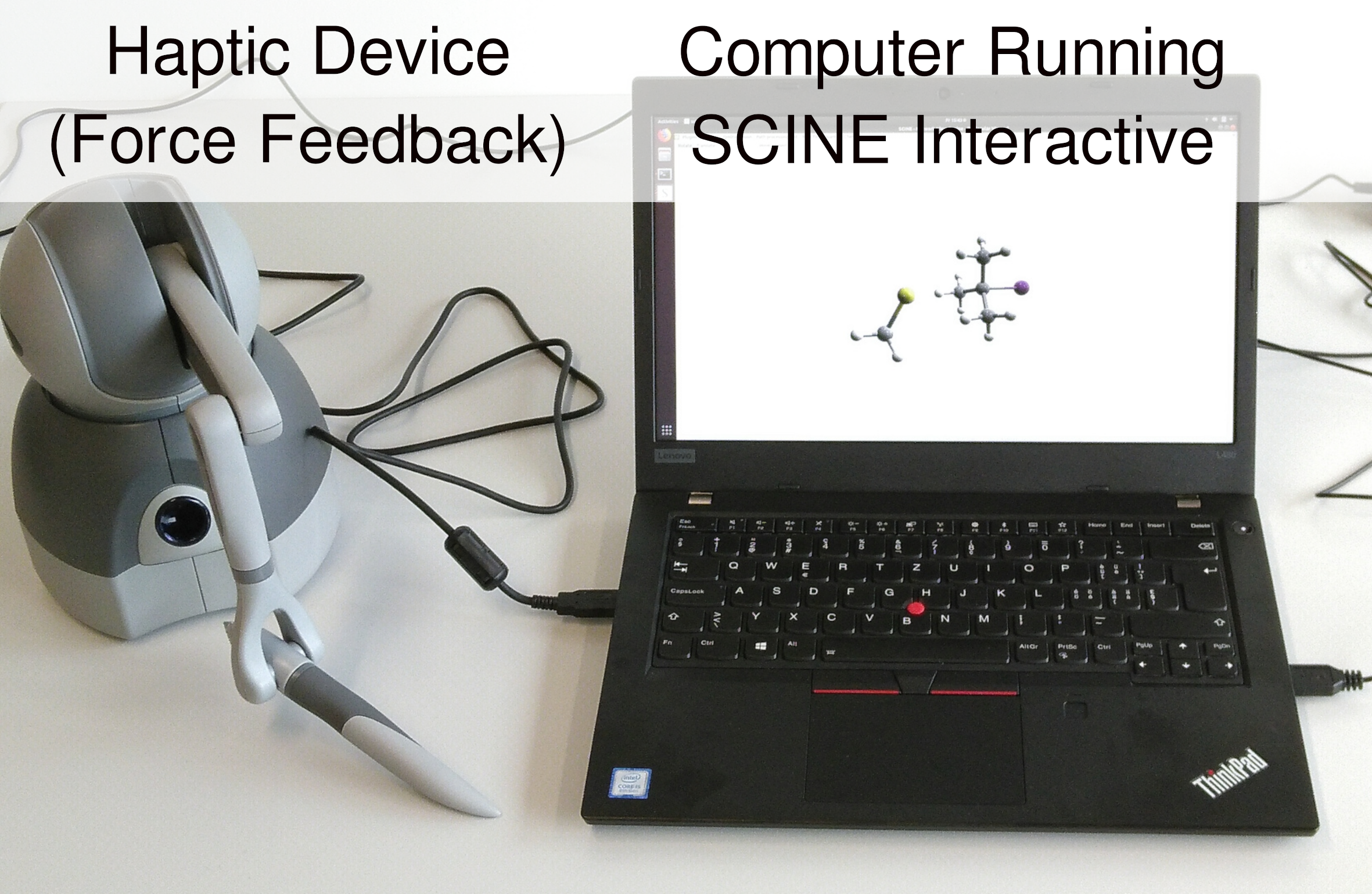}
\end{center}
\caption{\label{fig:haptic_device}\small A haptic pointer device (``Touch'' from 3D Systems) on the left (with 
the haptic pointer being in front) connected to a laptop on the right, set up to interactively explore a 
chemical reaction.}
\end{figure}

Another way to literally create a physical experience and to enhance the human perception of the molecular 
world is the introduction of haptic devices. A haptic device is a force-feedback hardware device, e.g., one that can
transmit a force through electric motors to some hardware attached to them (complex 
settings may even require one to wear an exoskeleton). Simplest variants that are commercially available 
comprise pen-like haptic pointers (see Fig.~\ref{fig:haptic_device}), which can be moved, rotated, and tilted 
freely in space. Motors in the joints of the arm of the haptic device can exert a force on the haptic pointer, 
which is then felt by the person operating the pointer. This force feedback addresses the haptic sense 
(\textit{i.e.}, the sense of touch) of the person, thereby complementing the visual and auditory senses 
typically engaged when working with virtual reality such as a molecule displayed on a computer screen. This 
renders the virtual reality more immersive and improves the intuitiveness with which one can interact 
with the molecular world. For example, a haptic device allows a one to experience the forces acting on a certain atom 
within a molecule in a very direct and intuitive way\cite{Marti2009}. Not surprisingly, previous research has 
demonstrated that haptic devices can be useful in chemical education\cite{minogue2006, Bivall2011, 
Zacharia2015, edwards2019}. Besides this use in chemical education and research, haptic devices are also 
employed, for example, in healthcare for surgical training\cite{lin2008, malone2010, ruikar2018}, and in the 
graphics industry for three-dimensional modeling\cite{dachille1999, lin2008}.

Irrespective of the way a student or researcher interacts with a virtual chemical system, it needs to behave according to 
the laws of quantum mechanics which govern the interaction of electrons and nuclei and their dynamics that give 
rise to its reactive behavior. Only when simulated in a physically reliable way, 
which guarantees a faithful and accurate representation of the molecular process, 
the virtual experience will 
be useful for educational (and research) purposes. Otherwise, students will not be able to reproduce and 
understand molecular processes as they would be plagued by artefacts and misleading experiences.
For a truly immersive learning experience, the physical simulation has to be done in real-time, such that the 
user is able to manipulate a molecule and experience the consequences of this manipulation without any 
noticeable delay. This is particularly challenging for a procedure rooted in quantum mechanics (and is the reason why
haptics and virtual reality was first explored with classical force fields instead).
Due to the ever-increasing performance of modern computers, interactive, real-time quantum 
chemistry has recently become possible\cite{Marti2009, bosson2012, Haag2013, Haag2014, Haag2014a, 
luehr2015, Vaucher2016}. The algorithmic developments made by our research group to accommodate interactive quantum 
chemistry are assembled in our software package called \textsc{SCINE Interactive}\cite{scine_interactive} as 
well as in the program package \textsc{Samson}\cite{samson}.

This paper demonstrates how chemical concepts can be understood by using haptic devices and interactive 
quantum chemistry. It is organized as follows: First, we briefly review interactive quantum chemistry and 
the \textsc{SCINE Interactive} software package, which is available free of charge on our webpage, and 
highlight their potential role in chemical education. Then, in 
section~\ref{sec:tasks}, we illustrate the usefulness of this new type of computer-assisted learning by 
means of a haptic device and real-time quantum simulations at the example of a few prototypical learning 
tasks. Finally, we provide conclusions and an outlook in section~\ref{sec:conclusion}.

\section{The \textsc{SCINE Interactive} Software Package in Chemical Education}
\label{sec:scine}

\textsc{SCINE Interactive}\cite{scine_interactive} is the original implementation of real-time quantum 
chemistry, a concept invented by us in 2013\cite{Haag2013}, for molecular structure and reactivity exploration
with self-consistent orbital optimizations. 
At the heart of real-time quantum 
chemistry are ultra-fast electronic structure calculations which deliver quantum chemical results (almost) 
instantaneously. This then enables a person to explore the potential energy surface of a chemical system in 
real time, immersively and interactively. The person can manipulate the molecular structure with a mouse or a 
haptic device and directly perceive the response of the system to this manipulation through visual and also 
haptic feedback.

For a truly interactive experience, visual feedback needs to be provided at a rate of about 60\,Hz, while 
haptic feedback requires a much higher rate of about 1000\,Hz\cite{Vaucher2016}. Currently, semi-empirical 
methods are the preferred approach to generate sufficiently reliable energies and 
forces on a timescale of a few to a few hundred milliseconds. Naturally, real-time quantum chemistry required the 
development of techniques accelerating electronic structure calculations\cite{Haag2013, Haag2014a, 
Muhlbach2016}. In particular, for the very high refresh rates needed for a haptic device, we introduced a 
mediator potential which locally approximates the true potential energy surface and which can be evaluated 
very efficiently\cite{Vaucher2016}. By virtue of these developments, the computing hardware necessary for an 
immersive interactive experience is far from being demanding. In fact, a standard laptop is sufficient 
(\textit{cf.}, Fig.~\ref{fig:haptic_device}), even if a haptic device is to be employed.

Being able to explore and manipulate a chemical system interactively and in real time allows one to understand 
many complicated concepts in an intuitive way. Next, we give specific examples how real-time quantum 
chemistry, possibly enhanced with a haptic device or a virtual reality framework, can be used for teaching chemistry.

\section{Example Tasks}
\label{sec:tasks}

\subsection{First Steps}

First, the students are given the Cartesian coordinates of the three nuclei of a squeezed water molecule 
as an XYZ file. The students should then recognize that with these coordinates, the H--O--H bond 
angle is 90$^{\circ}$, which is not the equilibrium structure of water. Upon loading these coordinates  
into \textsc{SCINE Interactive}, the students see that the structure promptly relaxes, due to the ultra-fast
quantum structure optimization running in the background, by opening up the H-O-H 
angle until the equilibrium structure is reached.

Another aspect which is slightly more subtle for most students to understand is that the XYZ file contains no 
information whatsoever about the bonds present in the molecule. Instead, this information is obtained from the 
quantum mechanical calculations running in the background of the interactive exploration. In fact, bonds are 
drawn in the graphical user interface based on a simple distance-based measure (that can be supplemented with 
bond order information taken from the underlying electronic wave function by drawing tubes between atomic 
spheres of increasing diameter with increasing bond order).

Students not having any prior experience operating a haptic device can use this simple example system to get 
acquainted with the interactive exploration of a chemical system with force feedback. In particular, they can 
develop a feeling for how strong a force they apply to an atom can be before the atom gets abstracted from the molecule 
(note, however, that there is a scaling factor that mediates between the molecular and the macroscopic force): 
As long as the force is below some value, it will not be possible to abstract the atom from the rest of the molecule; rather, 
upon a slight displacement of an atom, the other atoms will follow due to a continuously running structure 
optimization in the background that removes the excess energy in the system, and in effect the entire molecule 
translates in space. Abstraction of an atom will only be possible if a sufficiently strong force is applied. 
Once an atom has been abstracted, students will quickly discover that the system has a strong tendency to 
reassemble to a full water molecule. Only if the abstracted atom is quickly brought to a rather large distance 
(say, about ten \r{A}ngstroms), one finds a quasi-stable system consisting of an OH radical and a hydrogen 
atom. From such a situation, one can try to attach the separated hydrogen atom to the other hydrogen atom, 
rather than to the oxygen atom, effectively forming H-H-O. As it turns out, it is virtually impossible to 
create such a molecule (because the uncerlying quantum description prevents it and the haptic device allows one to
experience and thus learn this fact easily). Whenever one approaches the hydrogen atom of the OH group, the hydroxy radical rotates 
and an O-H bond is established. Also, when one very carefully approaches the OH group along the O-H bond, 
any slight (and, in practice, unavoidable) deviation from a perfect linear approach will induce a rotation of 
the hydroxy radical because of the continuously running structure optimization.

\subsection{Nucleophilic Substitution}

Another instructive example is the prototypical S\textsubscript{N}2 reaction of methanethiolate and 2-iodo-2-methyl propane:
\begin{equation}
\mathrm{
CH_3S^- + (CH_3)_3CI \rightarrow (CH_3)_3CSCH_3 + I^-~.
}
\end{equation}
When interactively exploring this reaction, students will quickly find that the reaction occurs readily; it is 
straightforward to carry out with a haptic device. In fact, without bothering too much about the relative 
orientation of the two molecules, as soon as the sulfur atom is brought close enough to the tertiary carbon 
atom of 2-iodo-2-methyl propane, the reaction occurs. One observes the typical pentagonal-bipyramidal 
transition state, \textit{i.e.}, for a short amount of time, the sulfur atom has already developed a partial 
bond to the tertiary carbon atom, while the iodine atom is still partially bound to the molecule.

For this reaction, the overall charge set in the quantum chemical calculations is $-$1. It is very instructive 
to study what happens when the system is considered overall electrically neutral. Even though one has removed 
only a single electron from the total system, one finds that the reaction is now almost impossible to induce. This 
is an excellent exercise to understand the role of a good nucleophile and a good leaving group. Only with a 
very careful, time-consuming operation, the desired end product can be created eventually. It is, however, 
much more likely that a completely different reaction altogether will be found first, in particular if one is 
not (yet) so skilled in the operation of a haptic device.

\begin{figure}[H]
\begin{center}
\includegraphics[scale=1.0]{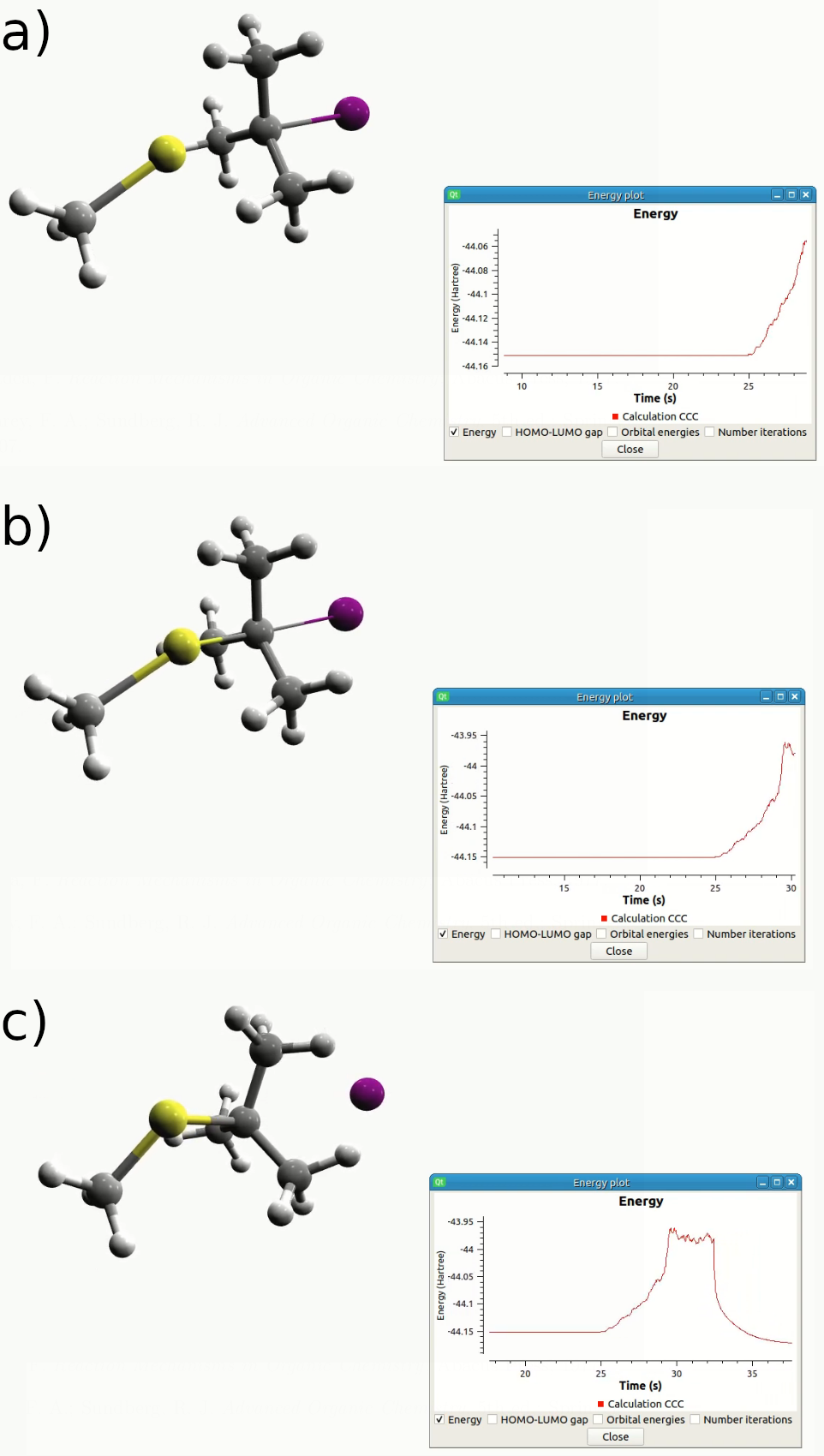}
\end{center}
\caption{\label{fig:sn2_reaction}\small Consecutive screenshots of the example S\textsubscript{N}2 reaction. 
a) The operator moves the  methanethiolate molecule towards the reactive center; note how the energy rises. b) 
Depiction of the approximate transition state. c) The bond to the I$^-$ leaving group is fully broken and the 
system is allowed to relax; note the energy set free in this process. The full movie is available on 
YouTube\cite{sn2_movie}.}
\end{figure}

Hence, one immediately understands that CH\textsubscript{3}S$^{-}$ is a good nucleophile, while 
CH\textsubscript{3}S is not (naturally, to create this insight for a novice requires 
a lot of additional explanation, while the real-time exploration transmits it as an immediate
puzzle whose explanation can then directly follow within the same setting -- by further visualizations such as charge density flow
during reaction, energy changes, and so forth). 
Likewise, I$^{-}$ is a good leaving group, 
while the neutral iodine atom is not. Using the haptic device, students can experience and literally feel the 
difference that a single electron can make in almost no time. A movie showcasing this example task can be found on 
YouTube\cite{sn2_movie}.

\subsection{Electrophilic Aromatic Substitution}

As a last example we consider the interactive exploration of the reaction between aniline and 
Br\textsuperscript{+}, \textit{i.e.}, an electrophilic aromatic substitution. We note in passing that also for 
this reaction care has to be taken to set the correct overall charge. Otherwise, spurious effects such as the 
ones described in the preceding section occur. When moving the bromine cation close to the benzene ring, 
students will find that both the \textit{ortho}- and the \textit{para}-substituted derivative can be created 
easily, and both are energetically strongly favored; about 230\,kJ/mol are released in both cases (which may 
be visualized by a color change of the background). Creating the \textit{meta}-substituted derivative is also  
possible without too much difficulty. However, the energy released is only about 40\,kJ/mol. Students can, 
therefore, directly witness the stereoelectronic effect of a directing group at work. Of course, it is 
straightforward to investigate the effect of different directing groups and even the combination of them. For 
example, by replacing the amino group by a methyl substituent, students will find that the energy released upon 
formation of the \textit{para}- and \textit{meta}-substituted derivatives are about 170\,kJ/mol and 130\,kJ/
mol, respectively. Hence, it is immediately obvious that the CH\textsubscript{3} group is a weaker 
\textit{ortho}- and \textit{para}-director compared to NH\textsubscript{2}.

This example is particularly interesting since the students will find that the creation of the $\sigma$-
complex from the (active) electrophile and the aromatic system is thermodynamically favored. In most textbooks 
of organic chemistry, this step is usually presented to be unfavored (see, \textit{e.g.}, 
Refs.~\citenum{vollhardt2003, sykes1986, badea1977}), even tough there are counterexamples (see, 
\textit{e.g.}, Ref.~\citenum{carey2007}).

The students will also find that upon adding the bromine cation to a carbon atom of the benzene ring, the bond 
between this carbon atom and the hydrogen atom will not break. Instead, the reaction will stop at the 
$\sigma$-complex. This clearly shows that the complete electrophilic aromatic substitution consists of two 
elementary steps. This stands in stark contrast to the previous example of an S\textsubscript{N}2 reaction, 
which features a single elementary step; the formation of a bond to the attacking group and the breaking of 
the bond to the leaving group happen in a concerted fashion. In an immersive interactive setting, students 
are able to experience, learn, and understand such concepts in an intuitive and descriptive way. A movie 
demonstrating this task can also be found on YouTube\cite{sear_movie}.

\begin{figure}[H]
\begin{center}
\includegraphics[scale=1.0]{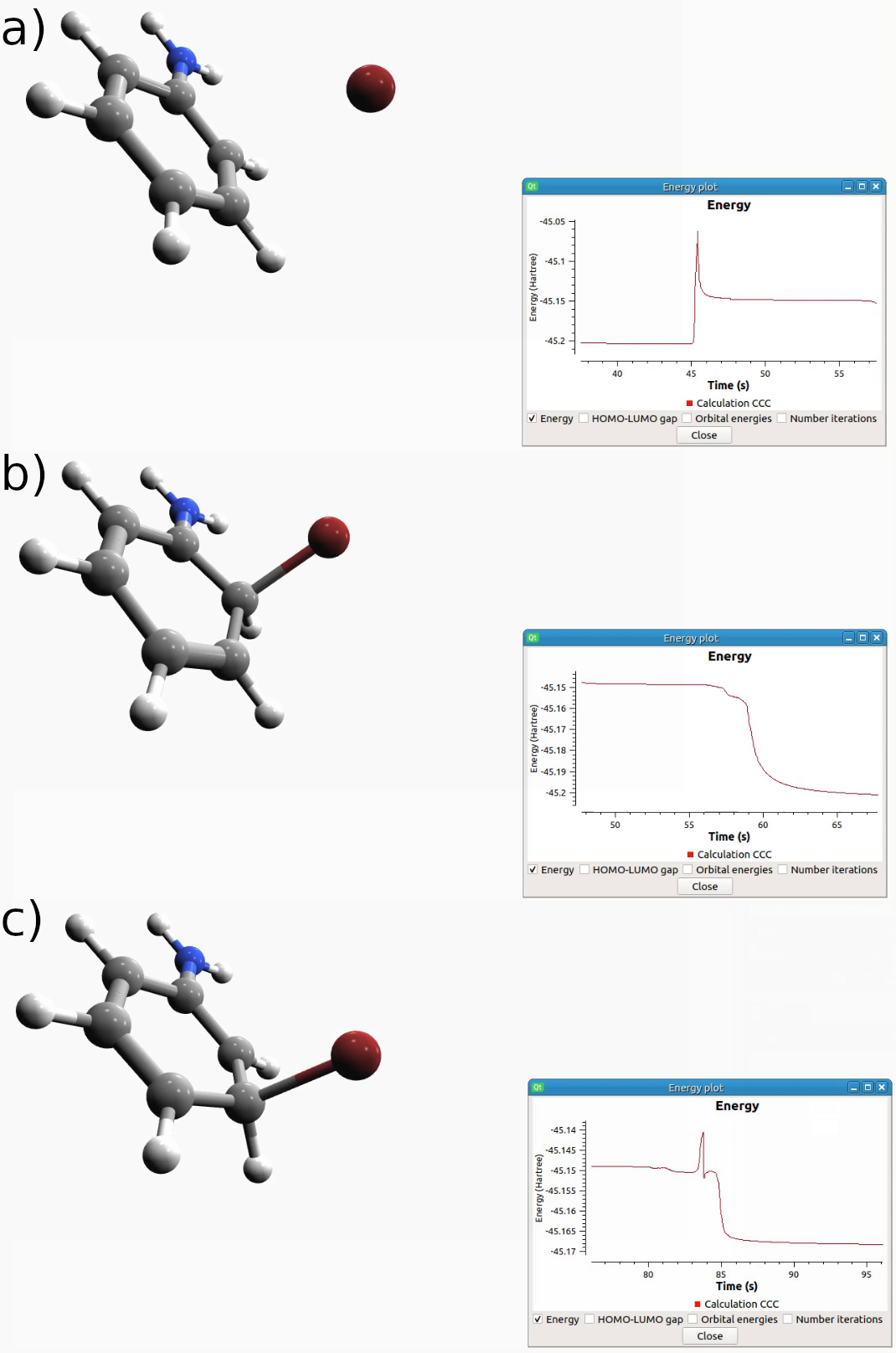}
\end{center}
\caption{\label{fig:ear_reaction}\small Consecutive screenshots of the example electrophilic aromatic 
substitution reaction. a) The operator approaches the benzene ring with the Br$^+$ cation. b) The 
\textit{ortho}-subsituted product is formed. c) The \textit{meta}-substituted product is formed. The full 
movie is available on YouTube\cite{sear_movie}.}
\end{figure}

\section{Conclusions and Outlook}
\label{sec:conclusion}

Modern approaches that rely on new computer hardware such as haptic (force-feedback) devices and virtual or 
augmented reality frameworks allow for immersive and hence more intuitive learning  experiences. In this 
article, we showcased some example learning tasks typical for an undergraduate chemistry curriculum. In 
particular, we demonstrated how an embodied-learning approach, making use of an interactive exploration and 
manipulation of the system by means of a haptic device, provides an illustrative and intuitive learning 
experience. This interactive exploration of the system is only possible by virtue of the advances made during the 
past years in the field of real-time quantum chemistry \cite{Haag2013}.

However, we should emphasize that the present setting focuses on the electronic contribution to reactions and reaction energies. In other
words, all reactions or chemical concepts dominated by electronic effects can be studied with our methodology. To also include
entropic (rather than these enthalpic) contributions will require further developments in the real-time quantum chemistry
framework (but such extensions are, in fact, already exploited 
in steered molecular dynamics settings\cite{jensen2002, stone2001} which exploit Jarzynski's identity\cite{jarzynski1997}).

It can be expected that such a learning setting can significantly improve the overall learning outcome. 
However, as a next step this needs to be rigorously investigated in extended users studies. Such work is currently 
being carried out in our laboratory within the Future Learning Initiative of ETH Zurich in collaboration with 
Prof.~Manu Kapur and his group.

\section*{Acknowledgments}
This work has been presented in a lecture with practical exercises for chemistry teachers at the fall meeting of the Swiss 
Chemical Society at the University of Zurich in September 2019.


\begin{mcitethebibliography}{61}
\providecommand*\natexlab[1]{#1}
\providecommand*\mciteSetBstSublistMode[1]{}
\providecommand*\mciteSetBstMaxWidthForm[2]{}
\providecommand*\mciteBstWouldAddEndPuncttrue
  {\def\EndOfBibitem{\unskip.}}
\providecommand*\mciteBstWouldAddEndPunctfalse
  {\let\EndOfBibitem\relax}
\providecommand*\mciteSetBstMidEndSepPunct[3]{}
\providecommand*\mciteSetBstSublistLabelBeginEnd[3]{}
\providecommand*\EndOfBibitem{}
\mciteSetBstSublistMode{f}
\mciteSetBstMaxWidthForm{subitem}{(\alph{mcitesubitemcount})}
\mciteSetBstSublistLabelBeginEnd
  {\mcitemaxwidthsubitemform\space}
  {\relax}
  {\relax}

\bibitem[H{\"{o}}ffler and Leutner(2007)H{\"{o}}ffler, and
  Leutner]{Hoffler2007}
H{\"{o}}ffler,~T.~N.; Leutner,~D. {Instructional animation versus static
  pictures: A meta-analysis}. \emph{Learn. Instr.} \textbf{2007}, \emph{17},
  722--738\relax
\mciteBstWouldAddEndPuncttrue
\mciteSetBstMidEndSepPunct{\mcitedefaultmidpunct}
{\mcitedefaultendpunct}{\mcitedefaultseppunct}\relax
\EndOfBibitem
\bibitem[Tversky \latin{et~al.}(2002)Tversky, Morrison, and
  Betrancourt]{Tversky2002}
Tversky,~B.; Morrison,~J.~B.; Betrancourt,~M. {Animation: can it facilitate?}
  \emph{Int. J. Hum. Comput. Stud.} \textbf{2002}, \emph{57}, 247--262\relax
\mciteBstWouldAddEndPuncttrue
\mciteSetBstMidEndSepPunct{\mcitedefaultmidpunct}
{\mcitedefaultendpunct}{\mcitedefaultseppunct}\relax
\EndOfBibitem
\bibitem[de~Koning and Tabbers(2011)de~Koning, and Tabbers]{DeKoning2011}
de~Koning,~B.~B.; Tabbers,~H.~K. {Facilitating Understanding of Movements in
  Dynamic Visualizations: an Embodied Perspective}. \emph{Educ. Psychol. Rev.}
  \textbf{2011}, \emph{23}, 501--521\relax
\mciteBstWouldAddEndPuncttrue
\mciteSetBstMidEndSepPunct{\mcitedefaultmidpunct}
{\mcitedefaultendpunct}{\mcitedefaultseppunct}\relax
\EndOfBibitem
\bibitem[van Gog \latin{et~al.}(2009)van Gog, Paas, Marcus, Ayres, and
  Sweller]{VanGog2009}
van Gog,~T.; Paas,~F.; Marcus,~N.; Ayres,~P.; Sweller,~J. {The Mirror Neuron
  System and Observational Learning: Implications for the Effectiveness of
  Dynamic Visualizations}. \emph{Educ. Psychol. Rev.} \textbf{2009}, \emph{21},
  21--30\relax
\mciteBstWouldAddEndPuncttrue
\mciteSetBstMidEndSepPunct{\mcitedefaultmidpunct}
{\mcitedefaultendpunct}{\mcitedefaultseppunct}\relax
\EndOfBibitem
\bibitem[Atkinson \latin{et~al.}(1977)Atkinson, Bond, Tribble, and
  Wilson]{atkinson1977}
Atkinson,~W.~D.; Bond,~K.~E.; Tribble,~G.~L.,~III; Wilson,~K.~R. {Computing
  with Feeling}. \emph{Comput. and Graphics} \textbf{1977}, \emph{2},
  97--103\relax
\mciteBstWouldAddEndPuncttrue
\mciteSetBstMidEndSepPunct{\mcitedefaultmidpunct}
{\mcitedefaultendpunct}{\mcitedefaultseppunct}\relax
\EndOfBibitem
\bibitem[Ouh-young \latin{et~al.}(1989)Ouh-young, Beard, and
  Brooks]{ouhyoung1989}
Ouh-young,~M.; Beard,~D.~V.; Brooks,~F.~P.,~Jr. {Force Display Performs Better
  than Visual Display in a Simple 6-D Docking Task}. {Proceedings, 1989
  International Conference on Robotics and Automation}. 1989; pp
  1462--1466\relax
\mciteBstWouldAddEndPuncttrue
\mciteSetBstMidEndSepPunct{\mcitedefaultmidpunct}
{\mcitedefaultendpunct}{\mcitedefaultseppunct}\relax
\EndOfBibitem
\bibitem[Surles \latin{et~al.}(1994)Surles, Richardson, Richardson, and
  Brooks]{surles1994}
Surles,~M.~C.; Richardson,~J.~S.; Richardson,~D.~C.; Brooks,~F.~P.,~Jr.
  {Sculpting proteins interactively: Continual energy minimization embedded in
  a graphical modeling system}. \emph{Prot. Sci.} \textbf{1994}, \emph{3},
  198--210\relax
\mciteBstWouldAddEndPuncttrue
\mciteSetBstMidEndSepPunct{\mcitedefaultmidpunct}
{\mcitedefaultendpunct}{\mcitedefaultseppunct}\relax
\EndOfBibitem
\bibitem[Leech \latin{et~al.}(1996)Leech, Prins, and Hermans]{leech1996}
Leech,~J.; Prins,~J.~F.; Hermans,~J. {SMD: Visual Steering of Molecular
  Dynamics for Protein Design}. \emph{IEEE Comput. Sci. Eng.} \textbf{1996},
  \emph{3}, 38--45\relax
\mciteBstWouldAddEndPuncttrue
\mciteSetBstMidEndSepPunct{\mcitedefaultmidpunct}
{\mcitedefaultendpunct}{\mcitedefaultseppunct}\relax
\EndOfBibitem
\bibitem[Stone \latin{et~al.}(2001)Stone, Gullingsrud, and Schulten]{stone2001}
Stone,~J.~E.; Gullingsrud,~J.; Schulten,~K. {A System for Interactive Molecular
  Dynamics Simulation}. {Proceedings of the 2001 Symposium on Interactive 3D
  Graphics}. 2001; pp 191--194\relax
\mciteBstWouldAddEndPuncttrue
\mciteSetBstMidEndSepPunct{\mcitedefaultmidpunct}
{\mcitedefaultendpunct}{\mcitedefaultseppunct}\relax
\EndOfBibitem
\bibitem[Gillet \latin{et~al.}(2005)Gillet, Sanner, Stoffler, and
  Olson]{gillet2005}
Gillet,~A.; Sanner,~M.; Stoffler,~D.; Olson,~A. {Tangible Interfaces for
  Structural Molecular Biology}. \emph{Structure} \textbf{2005}, \emph{13},
  483--491\relax
\mciteBstWouldAddEndPuncttrue
\mciteSetBstMidEndSepPunct{\mcitedefaultmidpunct}
{\mcitedefaultendpunct}{\mcitedefaultseppunct}\relax
\EndOfBibitem
\bibitem[Murayama \latin{et~al.}(2007)Murayama, Shimizu, Nam, Satoh, and
  Sato]{murayama2007}
Murayama,~J.; Shimizu,~H.; Nam,~C.~S.; Satoh,~H.; Sato,~M. {An Educational
  Environment for Chemical Contents with Haptic Interaction}. {2007
  International Conference on Cyberworlds (CW'07)}. 2007; pp 346--352\relax
\mciteBstWouldAddEndPuncttrue
\mciteSetBstMidEndSepPunct{\mcitedefaultmidpunct}
{\mcitedefaultendpunct}{\mcitedefaultseppunct}\relax
\EndOfBibitem
\bibitem[Limniou \latin{et~al.}(2008)Limniou, Roberts, and
  Papadopoulos]{limniou2008}
Limniou,~M.; Roberts,~D.; Papadopoulos,~N. {Full immersive virtual environment
  CAVE\textsuperscript{TM} in chemistry education}. \emph{Comput. Educ.}
  \textbf{2008}, \emph{51}, 584--593\relax
\mciteBstWouldAddEndPuncttrue
\mciteSetBstMidEndSepPunct{\mcitedefaultmidpunct}
{\mcitedefaultendpunct}{\mcitedefaultseppunct}\relax
\EndOfBibitem
\bibitem[Marti and Reiher(2009)Marti, and Reiher]{Marti2009}
Marti,~K.~H.; Reiher,~M. {Haptic Quantum Chemistry}. \emph{J. Comput. Chem.}
  \textbf{2009}, \emph{30}, 2010--2020\relax
\mciteBstWouldAddEndPuncttrue
\mciteSetBstMidEndSepPunct{\mcitedefaultmidpunct}
{\mcitedefaultendpunct}{\mcitedefaultseppunct}\relax
\EndOfBibitem
\bibitem[Delalande \latin{et~al.}(2009)Delalande, F\'{e}rey, Grasseau, and
  Baaden]{delalande2009}
Delalande,~O.; F\'{e}rey,~N.; Grasseau,~G.; Baaden,~M. {Complex Molecular
  Assemblies at Hand via Interactive Simulations}. \emph{J. Comput. Chem.}
  \textbf{2009}, \emph{30}, 2375--2387\relax
\mciteBstWouldAddEndPuncttrue
\mciteSetBstMidEndSepPunct{\mcitedefaultmidpunct}
{\mcitedefaultendpunct}{\mcitedefaultseppunct}\relax
\EndOfBibitem
\bibitem[Bosson \latin{et~al.}(2012)Bosson, Richard, Plet, Grudinin, and
  Redon]{bosson2012}
Bosson,~M.; Richard,~C.; Plet,~A.; Grudinin,~S.; Redon,~S. {Interactive Quantum
  Chemistry: A Divide‐and‐Conquer ASED‐MO Method}. \emph{J. Comput.
  Chem.} \textbf{2012}, \emph{33}, 779--790\relax
\mciteBstWouldAddEndPuncttrue
\mciteSetBstMidEndSepPunct{\mcitedefaultmidpunct}
{\mcitedefaultendpunct}{\mcitedefaultseppunct}\relax
\EndOfBibitem
\bibitem[Haag and Reiher(2013)Haag, and Reiher]{Haag2013}
Haag,~M.~P.; Reiher,~M. {Real-Time Quantum Chemistry}. \emph{Int. J. Quantum
  Chem.} \textbf{2013}, \emph{113}, 8--20\relax
\mciteBstWouldAddEndPuncttrue
\mciteSetBstMidEndSepPunct{\mcitedefaultmidpunct}
{\mcitedefaultendpunct}{\mcitedefaultseppunct}\relax
\EndOfBibitem
\bibitem[Haag and Reiher(2014)Haag, and Reiher]{Haag2014}
Haag,~M.~P.; Reiher,~M. {Studying chemical reactivity in a virtual
  environment}. \emph{Faraday Discuss.} \textbf{2014}, \emph{169},
  89--118\relax
\mciteBstWouldAddEndPuncttrue
\mciteSetBstMidEndSepPunct{\mcitedefaultmidpunct}
{\mcitedefaultendpunct}{\mcitedefaultseppunct}\relax
\EndOfBibitem
\bibitem[Haag \latin{et~al.}(2014)Haag, Vaucher, Bosson, Redon, and
  Reiher]{Haag2014a}
Haag,~M.~P.; Vaucher,~A.~C.; Bosson,~M.; Redon,~S.; Reiher,~M. {Interactive
  Chemical Reactivity Exploration}. \emph{ChemPhysChem} \textbf{2014},
  \emph{15}, 3301--3319\relax
\mciteBstWouldAddEndPuncttrue
\mciteSetBstMidEndSepPunct{\mcitedefaultmidpunct}
{\mcitedefaultendpunct}{\mcitedefaultseppunct}\relax
\EndOfBibitem
\bibitem[far(2014)]{faraday2014}
{Virtual and augmented reality immersive molecular simulations: general
  discussion}. \emph{Faraday Discuss.} \textbf{2014}, \emph{169},
  143--166\relax
\mciteBstWouldAddEndPuncttrue
\mciteSetBstMidEndSepPunct{\mcitedefaultmidpunct}
{\mcitedefaultendpunct}{\mcitedefaultseppunct}\relax
\EndOfBibitem
\bibitem[Glowacki \latin{et~al.}(2014)Glowacki, O'Connor, Calabr\'{o}, Price,
  Tew, Mitchell, Hyde, Tew, Coughtrie, and McIntosh-Smith]{glowacki2014}
Glowacki,~D.~R.; O'Connor,~M.; Calabr\'{o},~G.; Price,~J.; Tew,~P.;
  Mitchell,~T.; Hyde,~J.; Tew,~D.~P.; Coughtrie,~D.~J.; McIntosh-Smith,~S. {A
  GPU-accelerated immersive audio-visual framework for interaction with
  molecular dynamics using consumer depth sensors}. \emph{Faraday Discuss.}
  \textbf{2014}, \emph{169}, 63--87\relax
\mciteBstWouldAddEndPuncttrue
\mciteSetBstMidEndSepPunct{\mcitedefaultmidpunct}
{\mcitedefaultendpunct}{\mcitedefaultseppunct}\relax
\EndOfBibitem
\bibitem[Luehr \latin{et~al.}(2015)Luehr, Jin, and Mart\'{\i}nez]{luehr2015}
Luehr,~N.; Jin,~A. G.~B.; Mart\'{\i}nez,~T.~J. {\textit{Ab Initio} Interactive
  Molecular Dynamics on Graphical Processing Units (GPUs)}. \emph{J. Chem.
  Theory Comput.} \textbf{2015}, \emph{11}, 4536--4544\relax
\mciteBstWouldAddEndPuncttrue
\mciteSetBstMidEndSepPunct{\mcitedefaultmidpunct}
{\mcitedefaultendpunct}{\mcitedefaultseppunct}\relax
\EndOfBibitem
\bibitem[Norrby \latin{et~al.}(2015)Norrby, Grebner, Eriksson, and
  Bostr\"om]{norrby2015}
Norrby,~M.; Grebner,~C.; Eriksson,~J.; Bostr\"om,~J. {Molecular Rift: Virtual
  Reality for Drug Designers}. \emph{J. Chem. Inf. Model.} \textbf{2015},
  \emph{55}, 2475--2484\relax
\mciteBstWouldAddEndPuncttrue
\mciteSetBstMidEndSepPunct{\mcitedefaultmidpunct}
{\mcitedefaultendpunct}{\mcitedefaultseppunct}\relax
\EndOfBibitem
\bibitem[Salvadori \latin{et~al.}(2016)Salvadori, Frate, Pagliai, Mancini, and
  Barone]{salvadori2016}
Salvadori,~A.; Frate,~G.~D.; Pagliai,~M.; Mancini,~G.; Barone,~V. {Immersive
  virtual reality in computational chemistry: Applications to the analysis of
  QM and MM data}. \emph{Int. J. Quantum Chem.} \textbf{2016}, \emph{116},
  1731--1746\relax
\mciteBstWouldAddEndPuncttrue
\mciteSetBstMidEndSepPunct{\mcitedefaultmidpunct}
{\mcitedefaultendpunct}{\mcitedefaultseppunct}\relax
\EndOfBibitem
\bibitem[Zheng and Waller(2017)Zheng, and Waller]{zheng2017}
Zheng,~M.; Waller,~M.~P. {\textit{ChemPreview}: an augmented reality-based
  molecular interface}. \emph{J. Mol. Graphics Modell.} \textbf{2017},
  \emph{73}, 18--23\relax
\mciteBstWouldAddEndPuncttrue
\mciteSetBstMidEndSepPunct{\mcitedefaultmidpunct}
{\mcitedefaultendpunct}{\mcitedefaultseppunct}\relax
\EndOfBibitem
\bibitem[O'Connor \latin{et~al.}(4)O'Connor, Deeks, Dawn, Metatla, Roudaut,
  Sutton, Thomas, Glowacki, Sage, Tew, Wonnacott, Bates, Mulholland, and
  Glowacki]{oconnor2018}
O'Connor,~M.; Deeks,~H.~M.; Dawn,~E.; Metatla,~O.; Roudaut,~A.; Sutton,~M.;
  Thomas,~L.~M.; Glowacki,~B.~R.; Sage,~R.; Tew,~P. \latin{et~al.}  {Sampling
  molecular conformations and dynamics in a multiuser virtual reality
  framework}. \emph{Sci. Adv.} \textbf{4}, \emph{2018}, eaat2731\relax
\mciteBstWouldAddEndPuncttrue
\mciteSetBstMidEndSepPunct{\mcitedefaultmidpunct}
{\mcitedefaultendpunct}{\mcitedefaultseppunct}\relax
\EndOfBibitem
\bibitem[M\"uller \latin{et~al.}(2018)M\"uller, Krone, Huber, Biener, Herr,
  Koch, Reina, Weiskopf, and Ertl]{muller2018}
M\"uller,~C.; Krone,~M.; Huber,~M.; Biener,~V.; Herr,~D.; Koch,~S.; Reina,~G.;
  Weiskopf,~D.; Ertl,~T. {Interactive Molecular Graphics for Augmented Reality
  Using HoloLens}. \emph{J. Integr. Bioinform.} \textbf{2018}, \emph{15},
  20180005\relax
\mciteBstWouldAddEndPuncttrue
\mciteSetBstMidEndSepPunct{\mcitedefaultmidpunct}
{\mcitedefaultendpunct}{\mcitedefaultseppunct}\relax
\EndOfBibitem
\bibitem[Kingsley \latin{et~al.}(2019)Kingsley, Brunet, Lelais, McCloskey,
  Milliken, Leija, Fuhs, Wang, Zhou, and Spraggon]{kingsley2019}
Kingsley,~L.~J.; Brunet,~V.; Lelais,~G.; McCloskey,~S.; Milliken,~K.;
  Leija,~E.; Fuhs,~S.~R.; Wang,~K.; Zhou,~E.; Spraggon,~G. {Development of a
  virtual reality platform for effective communication of structural data in
  drug discovery}. \emph{J. Mol. Graphics Modell.} \textbf{2019}, \emph{89},
  234--241\relax
\mciteBstWouldAddEndPuncttrue
\mciteSetBstMidEndSepPunct{\mcitedefaultmidpunct}
{\mcitedefaultendpunct}{\mcitedefaultseppunct}\relax
\EndOfBibitem
\bibitem[Garc\'{i}a-Hern\'{a}ndez and
  Kranzm\"uller(2019)Garc\'{i}a-Hern\'{a}ndez, and
  Kranzm\"uller]{garciahernandez2019}
Garc\'{i}a-Hern\'{a}ndez,~R.~J.; Kranzm\"uller,~D. {NOMAD VR: Multiplatform
  virtual reality viewer for chemistry simulations}. \emph{Comput. Phys.
  Commun.} \textbf{2019}, \emph{237}, 230--237\relax
\mciteBstWouldAddEndPuncttrue
\mciteSetBstMidEndSepPunct{\mcitedefaultmidpunct}
{\mcitedefaultendpunct}{\mcitedefaultseppunct}\relax
\EndOfBibitem
\bibitem[Ferrell \latin{et~al.}(2019)Ferrell, Campbell, McCarthy, McKay,
  Hensinger, Srinivasa, Zhao, Wurthmann, Li, and Schneebeli]{ferrell2019}
Ferrell,~J.~B.; Campbell,~J.~P.; McCarthy,~D.~R.; McKay,~K.~T.; Hensinger,~M.;
  Srinivasa,~R.; Zhao,~X.; Wurthmann,~A.; Li,~J.; Schneebeli,~S.~T. {Chemical
  Exploration with Virtual Reality in Organic Teaching Laboratories}. \emph{J.
  Chem. Educ.} \textbf{2019}, \emph{96}, 1961--1966\relax
\mciteBstWouldAddEndPuncttrue
\mciteSetBstMidEndSepPunct{\mcitedefaultmidpunct}
{\mcitedefaultendpunct}{\mcitedefaultseppunct}\relax
\EndOfBibitem
\bibitem[Dai \latin{et~al.}(2020)Dai, Laureanti, Kopelevich, and
  Diaconescu]{dai2020}
Dai,~R.; Laureanti,~J.~A.; Kopelevich,~M.; Diaconescu,~P.~L. {Developing a
  Virtual Reality Approach toward a Better Understanding of Coordination
  Chemistry and Molecular Orbitals}. \emph{J. Chem. Educ.} \textbf{2020},
  \emph{97}, 3647--3651\relax
\mciteBstWouldAddEndPuncttrue
\mciteSetBstMidEndSepPunct{\mcitedefaultmidpunct}
{\mcitedefaultendpunct}{\mcitedefaultseppunct}\relax
\EndOfBibitem
\bibitem[Doak \latin{et~al.}(2020)Doak, Denyer, Gerrard, Mackay, and
  Allison]{doak2020}
Doak,~D.~G.; Denyer,~G.~S.; Gerrard,~J.~A.; Mackay,~J.~P.; Allison,~J.~R.
  {Peppy: A virtual reality environment for exploring the principles of
  polypeptide structure}. \emph{Prot. Sci.} \textbf{2020}, \emph{29},
  157--168\relax
\mciteBstWouldAddEndPuncttrue
\mciteSetBstMidEndSepPunct{\mcitedefaultmidpunct}
{\mcitedefaultendpunct}{\mcitedefaultseppunct}\relax
\EndOfBibitem
\bibitem[Gandhi \latin{et~al.}(2020)Gandhi, Jakymiw, Barrett, Mahaseth, and
  White]{gandhi2020}
Gandhi,~H.~A.; Jakymiw,~S.; Barrett,~R.; Mahaseth,~H.; White,~A.~D. {Real-Time
  Interactive Simulation and Visualization of Organic Molecules}. \emph{J.
  Chem. Educ.} \textbf{2020}, {DOI 10.1021/acs.jchemed.9b01161}\relax
\mciteBstWouldAddEndPuncttrue
\mciteSetBstMidEndSepPunct{\mcitedefaultmidpunct}
{\mcitedefaultendpunct}{\mcitedefaultseppunct}\relax
\EndOfBibitem
\bibitem[Sanii(2020)]{sanii2020}
Sanii,~B. {Creating Augmented Reality USDZ Files to Visualize 3D Objects on
  Student Phones in the Classroom}. \emph{J. Chem. Educ.} \textbf{2020},
  \emph{97}, 253--257\relax
\mciteBstWouldAddEndPuncttrue
\mciteSetBstMidEndSepPunct{\mcitedefaultmidpunct}
{\mcitedefaultendpunct}{\mcitedefaultseppunct}\relax
\EndOfBibitem
\bibitem[mol()]{molegram}
{https://cadd.ethz.ch/education/hololense.html (accessed November 04,
  2020)}\relax
\mciteBstWouldAddEndPuncttrue
\mciteSetBstMidEndSepPunct{\mcitedefaultmidpunct}
{\mcitedefaultendpunct}{\mcitedefaultseppunct}\relax
\EndOfBibitem
\bibitem[Dede \latin{et~al.}(1999)Dede, Salzman, Loftin, and Sprague]{dede1999}
Dede,~C.; Salzman,~M.~C.; Loftin,~R.~B.; Sprague,~D. {Multisensory Immersion as
  a Modeling Environment for Learning Complex Scientific Concepts}.
  \textbf{1999}, 282--319\relax
\mciteBstWouldAddEndPuncttrue
\mciteSetBstMidEndSepPunct{\mcitedefaultmidpunct}
{\mcitedefaultendpunct}{\mcitedefaultseppunct}\relax
\EndOfBibitem
\bibitem[Trindade \latin{et~al.}(2002)Trindade, Fiolhais, and
  Almeida]{trindade2002}
Trindade,~J.; Fiolhais,~C.; Almeida,~L. {Science learning in virtual
  environments: a descriptive study}. \emph{Br. J. Educ. Technol.}
  \textbf{2002}, \emph{33}, 471--488\relax
\mciteBstWouldAddEndPuncttrue
\mciteSetBstMidEndSepPunct{\mcitedefaultmidpunct}
{\mcitedefaultendpunct}{\mcitedefaultseppunct}\relax
\EndOfBibitem
\bibitem[Bennie \latin{et~al.}(2019)Bennie, Ranaghan, Deeks, Goldsmith,
  O'Connor, Mulholland, and Glowacki]{bennie2019}
Bennie,~S.~J.; Ranaghan,~K.~E.; Deeks,~H.; Goldsmith,~H.~E.; O'Connor,~M.~B.;
  Mulholland,~A.~J.; Glowacki,~D.~R. {Teaching Enzyme Catalysis Using
  Interactive Molecular Dynamics in Virtual Reality}. \emph{J. Chem. Educ.}
  \textbf{2019}, \emph{96}, 2488--2496\relax
\mciteBstWouldAddEndPuncttrue
\mciteSetBstMidEndSepPunct{\mcitedefaultmidpunct}
{\mcitedefaultendpunct}{\mcitedefaultseppunct}\relax
\EndOfBibitem
\bibitem[Fung \latin{et~al.}(2019)Fung, Choo, Ardisara, Zimmermann, Watts,
  Koscielniak, Blanc, Coumoul, and Dumke]{fung2019}
Fung,~F.~M.; Choo,~W.~Y.; Ardisara,~A.; Zimmermann,~C.~D.; Watts,~S.;
  Koscielniak,~T.; Blanc,~E.; Coumoul,~X.; Dumke,~R. {Applying a Virtual
  Reality Platform in Environmental Chemistry Education To Conduct a Field Trip
  to an Overseas Site}. \emph{J. Chem. Educ.} \textbf{2019}, \emph{96},
  382--386\relax
\mciteBstWouldAddEndPuncttrue
\mciteSetBstMidEndSepPunct{\mcitedefaultmidpunct}
{\mcitedefaultendpunct}{\mcitedefaultseppunct}\relax
\EndOfBibitem
\bibitem[Jim\'{e}nez(2019)]{jimenez2019}
Jim\'{e}nez,~Z.~A. \emph{{Technology Integration in Chemistry Education and
  Research (TICER)}}; American Chemical Society, 2019; pp 31--52\relax
\mciteBstWouldAddEndPuncttrue
\mciteSetBstMidEndSepPunct{\mcitedefaultmidpunct}
{\mcitedefaultendpunct}{\mcitedefaultseppunct}\relax
\EndOfBibitem
\bibitem[Ali and Ullah(2020)Ali, and Ullah]{ali2020}
Ali,~N.; Ullah,~S. {Review to Analyze and Compare Virtual Chemistry
  Laboratories for Their Use in Education}. \emph{J. Chem. Educ.}
  \textbf{2020}, \emph{97}, 3563--3574\relax
\mciteBstWouldAddEndPuncttrue
\mciteSetBstMidEndSepPunct{\mcitedefaultmidpunct}
{\mcitedefaultendpunct}{\mcitedefaultseppunct}\relax
\EndOfBibitem
\bibitem[Minogue and Jones(2006)Minogue, and Jones]{minogue2006}
Minogue,~J.; Jones,~M.~G. {Haptics in Education: Exploring an Untapped Sensory
  Modality}. \emph{Rev. Educ. Res.} \textbf{2006}, \emph{76}, 317--348\relax
\mciteBstWouldAddEndPuncttrue
\mciteSetBstMidEndSepPunct{\mcitedefaultmidpunct}
{\mcitedefaultendpunct}{\mcitedefaultseppunct}\relax
\EndOfBibitem
\bibitem[Bivall \latin{et~al.}(2011)Bivall, Ainsworth, and Tibell]{Bivall2011}
Bivall,~P.; Ainsworth,~S.; Tibell,~L. A.~E. {Do Haptic Representations Help
  Complex Molecular Learning?} \emph{Sci. Educ.} \textbf{2011}, \emph{95},
  700--719\relax
\mciteBstWouldAddEndPuncttrue
\mciteSetBstMidEndSepPunct{\mcitedefaultmidpunct}
{\mcitedefaultendpunct}{\mcitedefaultseppunct}\relax
\EndOfBibitem
\bibitem[Zacharia(2015)]{Zacharia2015}
Zacharia,~Z.~C. {Examining whether touch sensory feedback is necessary for
  science learning through experimentation: A literature review of two
  different lines of research across K-16}. \emph{Educ. Res. Rev.}
  \textbf{2015}, \emph{16}, 116--137\relax
\mciteBstWouldAddEndPuncttrue
\mciteSetBstMidEndSepPunct{\mcitedefaultmidpunct}
{\mcitedefaultendpunct}{\mcitedefaultseppunct}\relax
\EndOfBibitem
\bibitem[Edwards \latin{et~al.}(2019)Edwards, Bielawski, Prada, and
  Cheok]{edwards2019}
Edwards,~B.~I.; Bielawski,~K.~S.; Prada,~R.; Cheok,~A.~D. {Haptic virtual
  reality and immersive learning for enhanced organic chemistry instruction}.
  \emph{Virtual Real.} \textbf{2019}, \emph{23}, 363--373\relax
\mciteBstWouldAddEndPuncttrue
\mciteSetBstMidEndSepPunct{\mcitedefaultmidpunct}
{\mcitedefaultendpunct}{\mcitedefaultseppunct}\relax
\EndOfBibitem
\bibitem[Lin and Otaduy(2008)Lin, and Otaduy]{lin2008}
Lin,~M.~C., Otaduy,~M., Eds. \emph{{Haptic Rendering}}; CRC Press, 2008\relax
\mciteBstWouldAddEndPuncttrue
\mciteSetBstMidEndSepPunct{\mcitedefaultmidpunct}
{\mcitedefaultendpunct}{\mcitedefaultseppunct}\relax
\EndOfBibitem
\bibitem[Malone \latin{et~al.}(2010)Malone, Syed, Downes, D'Ambrosio, Quest,
  and Kaiser]{malone2010}
Malone,~H.~R.; Syed,~O.~N.; Downes,~M.~S.; D'Ambrosio,~A.~L.; Quest,~D.~O.;
  Kaiser,~M.~G. {Simulation in Neurosurgery: A Review of Computer-Based
  Simulation Environments and Their Surgical Applications}. \emph{Neurosurg.}
  \textbf{2010}, \emph{67}, 1105--1116\relax
\mciteBstWouldAddEndPuncttrue
\mciteSetBstMidEndSepPunct{\mcitedefaultmidpunct}
{\mcitedefaultendpunct}{\mcitedefaultseppunct}\relax
\EndOfBibitem
\bibitem[Ruikar \latin{et~al.}(2018)Ruikar, Hegadi, and Santosh]{ruikar2018}
Ruikar,~D.~D.; Hegadi,~R.~S.; Santosh,~K.~C. {A Systematic Review on Orthopedic
  Simulators for Psycho-Motor Skill and Surgical Procedure Training}. \emph{J.
  Med. Syst.} \textbf{2018}, \emph{42}, 168\relax
\mciteBstWouldAddEndPuncttrue
\mciteSetBstMidEndSepPunct{\mcitedefaultmidpunct}
{\mcitedefaultendpunct}{\mcitedefaultseppunct}\relax
\EndOfBibitem
\bibitem[Dachille \latin{et~al.}(1999)Dachille, Qin, Kaufman, and
  El-Sana]{dachille1999}
Dachille,~F.; Qin,~H.; Kaufman,~A.; El-Sana,~J. {Haptic Sculpting of Dynamic
  Surfaces}. {Proceedings of the 1999 Symposium on Interactive 3D Graphics}.
  1999; p 103–110\relax
\mciteBstWouldAddEndPuncttrue
\mciteSetBstMidEndSepPunct{\mcitedefaultmidpunct}
{\mcitedefaultendpunct}{\mcitedefaultseppunct}\relax
\EndOfBibitem
\bibitem[Vaucher \latin{et~al.}(2016)Vaucher, Haag, and Reiher]{Vaucher2016}
Vaucher,~A.~C.; Haag,~M.~P.; Reiher,~M. {Real-Time Feedback from Iterative
  Electronic Structure Calculations}. \emph{J. Comput. Chem.} \textbf{2016},
  \emph{37}, 805--812\relax
\mciteBstWouldAddEndPuncttrue
\mciteSetBstMidEndSepPunct{\mcitedefaultmidpunct}
{\mcitedefaultendpunct}{\mcitedefaultseppunct}\relax
\EndOfBibitem
\bibitem[sci()]{scine_interactive}
{https://scine.ethz.ch/download/interactive (accessed November 04, 2020)}\relax
\mciteBstWouldAddEndPuncttrue
\mciteSetBstMidEndSepPunct{\mcitedefaultmidpunct}
{\mcitedefaultendpunct}{\mcitedefaultseppunct}\relax
\EndOfBibitem
\bibitem[sam()]{samson}
{https://www.samson-connect.net (accessed November 04, 2020)}\relax
\mciteBstWouldAddEndPuncttrue
\mciteSetBstMidEndSepPunct{\mcitedefaultmidpunct}
{\mcitedefaultendpunct}{\mcitedefaultseppunct}\relax
\EndOfBibitem
\bibitem[M{\"{u}}hlbach \latin{et~al.}(2016)M{\"{u}}hlbach, Vaucher, and
  Reiher]{Muhlbach2016}
M{\"{u}}hlbach,~A.~H.; Vaucher,~A.~C.; Reiher,~M. {Accelerating Wave Function
  Convergence in Interactive Quantum Chemical Reactivity Studies}. \emph{J.
  Chem. Theory Comput.} \textbf{2016}, \emph{12}, 1228--1235\relax
\mciteBstWouldAddEndPuncttrue
\mciteSetBstMidEndSepPunct{\mcitedefaultmidpunct}
{\mcitedefaultendpunct}{\mcitedefaultseppunct}\relax
\EndOfBibitem
\bibitem[sn2()]{sn2_movie}
{https://youtu.be/imYmcWboj8o (accessed November 04, 2020)}\relax
\mciteBstWouldAddEndPuncttrue
\mciteSetBstMidEndSepPunct{\mcitedefaultmidpunct}
{\mcitedefaultendpunct}{\mcitedefaultseppunct}\relax
\EndOfBibitem
\bibitem[Vollhardt and Schore(2003)Vollhardt, and Schore]{vollhardt2003}
Vollhardt,~K. P.~C.; Schore,~N.~E. \emph{{Organic Chemistry: Structure and
  Function}}, 4th ed.; W. H. Freeman and Company, 2003\relax
\mciteBstWouldAddEndPuncttrue
\mciteSetBstMidEndSepPunct{\mcitedefaultmidpunct}
{\mcitedefaultendpunct}{\mcitedefaultseppunct}\relax
\EndOfBibitem
\bibitem[Sykes(1986)]{sykes1986}
Sykes,~P. \emph{{A Guidebook to Mechanism in Organic Chemistry}}, 6th ed.;
  Longman Group UK Ltd., 1986\relax
\mciteBstWouldAddEndPuncttrue
\mciteSetBstMidEndSepPunct{\mcitedefaultmidpunct}
{\mcitedefaultendpunct}{\mcitedefaultseppunct}\relax
\EndOfBibitem
\bibitem[Badea(1977)]{badea1977}
Badea,~F. \emph{{Reaction Mechanisms in Organic Chemistry}}; Abacus Press,
  1977\relax
\mciteBstWouldAddEndPuncttrue
\mciteSetBstMidEndSepPunct{\mcitedefaultmidpunct}
{\mcitedefaultendpunct}{\mcitedefaultseppunct}\relax
\EndOfBibitem
\bibitem[Carey and Sundberg(2007)Carey, and Sundberg]{carey2007}
Carey,~F.~A.; Sundberg,~R.~J. \emph{{Advanced Organic Chemistry}}, 5th ed.;
  Springer, 2007\relax
\mciteBstWouldAddEndPuncttrue
\mciteSetBstMidEndSepPunct{\mcitedefaultmidpunct}
{\mcitedefaultendpunct}{\mcitedefaultseppunct}\relax
\EndOfBibitem
\bibitem[sea()]{sear_movie}
{https://youtu.be/cnoodBFfcVM (accessed November 04, 2020)}\relax
\mciteBstWouldAddEndPuncttrue
\mciteSetBstMidEndSepPunct{\mcitedefaultmidpunct}
{\mcitedefaultendpunct}{\mcitedefaultseppunct}\relax
\EndOfBibitem
\bibitem[Jensen \latin{et~al.}(2002)Jensen, Park, Tajkhorshid, and
  Schulten]{jensen2002}
Jensen,~M.~O.; Park,~S.; Tajkhorshid,~E.; Schulten,~K. {Energetics of glycerol
  conduction through aquaglyceroporin GlpF}. \emph{Proc. Natl. Acad. Sci.
  U.S.A.} \textbf{2002}, \emph{99}, 6731--6736\relax
\mciteBstWouldAddEndPuncttrue
\mciteSetBstMidEndSepPunct{\mcitedefaultmidpunct}
{\mcitedefaultendpunct}{\mcitedefaultseppunct}\relax
\EndOfBibitem
\bibitem[Jarzynski(1997)]{jarzynski1997}
Jarzynski,~C. {Nonequilibrium Equality for Free Energy Differences}.
  \emph{Phys. Rev. Lett.} \textbf{1997}, \emph{78}, 2690--2693\relax
\mciteBstWouldAddEndPuncttrue
\mciteSetBstMidEndSepPunct{\mcitedefaultmidpunct}
{\mcitedefaultendpunct}{\mcitedefaultseppunct}\relax
\EndOfBibitem
\end{mcitethebibliography}
\providecommand{\latin}[1]{#1}
\makeatletter
\providecommand{\doi}
  {\begingroup\let\do\@makeother\dospecials
  \catcode`\{=1 \catcode`\}=2 \doi@aux}
\providecommand{\doi@aux}[1]{\endgroup\texttt{#1}}
\makeatother
\providecommand*\mcitethebibliography{\thebibliography}
\csname @ifundefined\endcsname{endmcitethebibliography}
  {\let\endmcitethebibliography\endthebibliography}{}

\end{document}